\documentclass[aps,prl,twocolumn,superscriptaddress,amsmath,amssymb,amsfonts,10pt]{revtex4-2}

\usepackage{graphicx}
\usepackage{braket}
\usepackage[dvipsnames]{xcolor}
\usepackage[normalem]{ulem}
\usepackage[bookmarks=false,colorlinks=true,urlcolor=RoyalBlue,citecolor=NavyBlue,linkcolor=OrangeRed]{hyperref}
\usepackage{bm,bbm}

\newcommand{\CNOT}{\ensuremath{\mathcal{C}\hspace{-.02cm}\mathcal{N}\hspace{-.06cm}\mathcal{O}\hspace{-.02cm}\mathcal{T}}}

\begin{document}

\title{Profusion of Symmetry-Protected Qubits from Stable Ergodicity Breaking}

\author{Thomas Iadecola}
\email{iadecola@psu.edu}
\affiliation{Department of Physics, The Pennsylvania State University, University Park, Pennsylvania 16802, USA}
\affiliation{Institute for Computational and Data Sciences, The Pennsylvania State University, University Park, Pennsylvania 16802, USA}
\affiliation{Materials Research Institute, The Pennsylvania State University, University Park, Pennsylvania 16802, USA}

\author{Rahul Nandkishore}
\email{rahul.nandkishore@colorado.edu}
\affiliation{Department of Physics and Center for Theory of Quantum Matter, University of Colorado Boulder, Boulder, Colorado 80309 USA}

\date{\today}

\begin{abstract}
We show how combining a discrete symmetry with topological Hilbert space fragmentation can give rise to exponentially many topologically stable qubits protected by a single discrete symmetry. We illustrate this explicitly with the example of the $\mathsf{CZ}_p$ model, where the encoded qubits are stable to arbitrary symmetry-respecting perturbations for parametrically long times, substantially enhancing the robustness of a recently proposed construction based on nontopological fragmentation. In this model, the encoded qubits naturally come in pairs for which a universal set of transversal logical gates can be performed, ruling out (by the Eastin-Knill theorem) the possibility of using them for quantum error correction. We also comment on the combination of symmetry enrichment and topological fragmentation more generally, and the implications for use of systems exhibiting Hilbert space fragmentation as quantum memories. 
\end{abstract}

\maketitle

Recent decades have witnessed a rapid evolution in our understanding of nonequilibrium quantum matter. While textbook treatments assume that generic many-particle systems should equilibrate under their own dynamics, we have come to realize that in many circumstances, many-particle quantum systems can fail to equilibrate under their own dynamics, violating the ergodic hypothesis. Routes to ``ergodicity breaking" of this type include integrability (see, e.g., \cite{Baxter}), many body localization~\cite{MBLARCMP, MBLRMP}, quantum many-body scars~\cite{Serbyn21,Moudgalya22a,Chandran23}, and Hilbert space shattering [aka {\it Hilbert space fragmentation} (HSF)]\cite{KHN, Sala20, Moudgalya2022Thermalization}. This last phenomenon, which will be the focus of this paper, arises from the interplay between judiciously chosen conservation laws or constraints. While originally discovered in the context of charge and dipole conserving dynamics~\cite{PPN}, it has since been found to arise in a range of settings, including the quantum Ising model \cite{Yoshinaga, HartIsing} and lattice gauge theories~\cite{Yang20}. It can be made topologically robust \cite{stephen2022ergodicity, SNH, KSHN, robuster}, and described with a set of powerful mathematical tools, including commutant algebras~\cite{moudgalya2022hilbert} and geometric group theory~\cite{balasubramanian2023glassy, KSHN}. 

There is an intimate connection between ergodicity-breaking dynamics and quantum memories. In particular, insofar as ergodicity breaking dynamics involve the preservation of a memory of the initial conditions in local observables for long times, they may seem natural candidates for (passive) quantum memories. However, until recently it has been believed that HSF constituted only a classical, not a quantum memory, since diagonal perturbations could always be added \cite{KHN} spoiling the phase information. This state of affairs was remedied in \cite{Iadecola25}, where one of the present authors pointed out that symmetry enrichment could enable HSF to serve as a route to quantum memory protected by symmetry and by the HSF structure. That work took as its primary example the XNOR model~\cite{Yang20,Singh21,Pozsgay21,Zadnik21a,Zadnik21b,Vuina25}, which exhibits garden-variety HSF with a fairly limited degree of robustness. Far more robust examples of HSF exist, including models exhibiting topological stability~\cite{stephen2022ergodicity, SNH, KSHN, robuster}. Can we combine the symmetry enrichment prescription of~\cite{Iadecola25} with topologically stable HSF to obtain topologically stable qubits protected by symmetry alone? 

In this manuscript we show how one may combine ideas of symmetry enrichment from \cite{Iadecola25} with topologically stable HSF to obtain models which encode an exponentially large number of stable qubits with topological stability, protected by a single discrete symmetry. We explicitly demonstrate this in the context of the $\mathsf{CZ}_p$ model, the first and simplest exemplar of topological HSF \cite{stephen2022ergodicity}, and comment on the generalization to other models of this type. We show that this model not only encodes exponentially many topologically stable qubits, but also that these qubits naturally come in pairs that allow for the transversal application of a universal set of quantum logic gates. It does not, however, constitute an error correcting code, in accordance with the no-go theorem of Eastin and Knill~\cite{Eastin09}, since it is unprotected against errors which violate the protecting symmetry. 

\begin{figure}[t!]
    \centering
    \includegraphics[width=0.8\columnwidth]{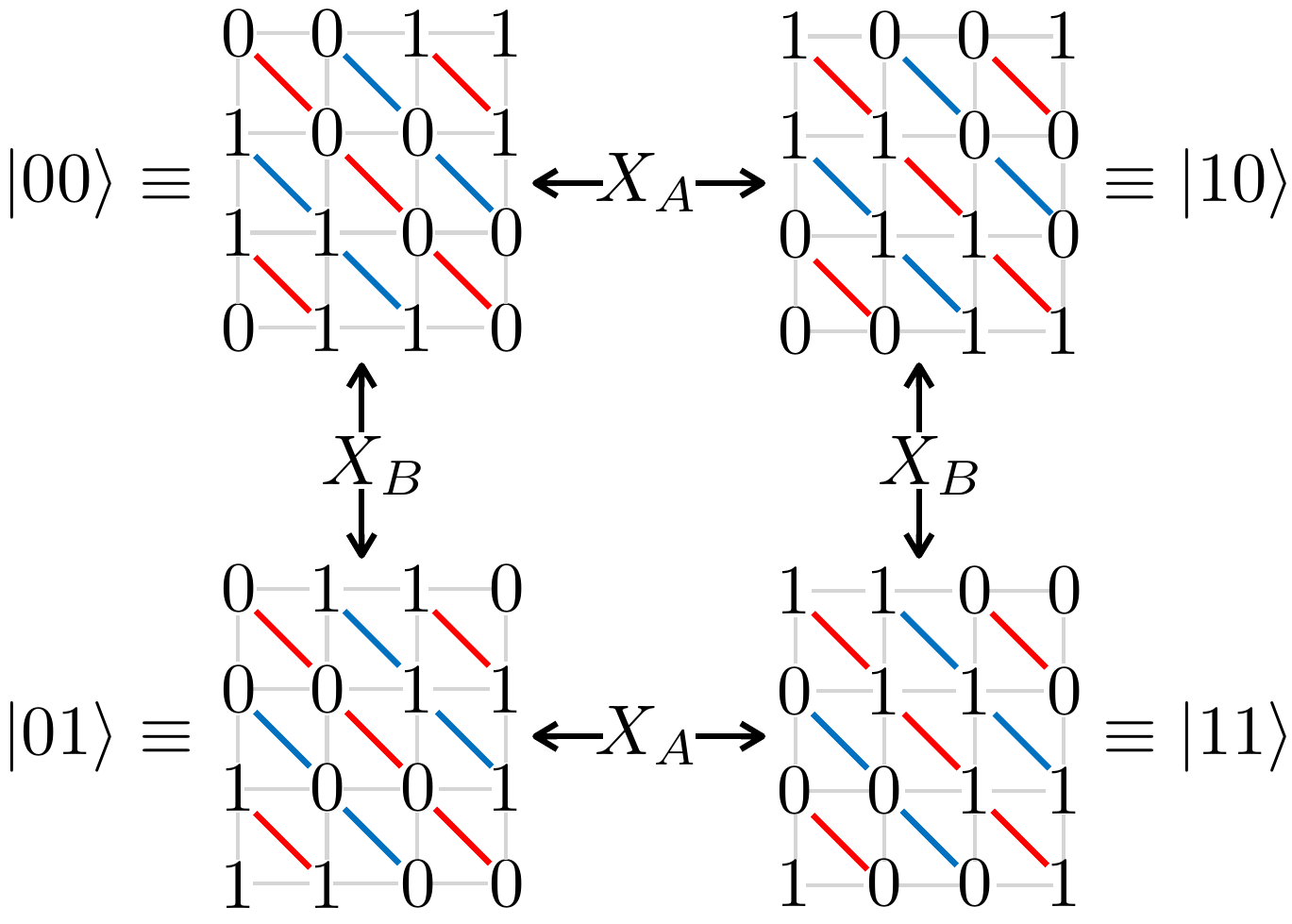}
    \caption{
    Schematic depiction of the two-qubit logical space spanned by a set of four symmetry-related frozen states. Blue and red lines represent domain walls on the $A$ and $B$ sublattices, respectively. Frozen states are connected by applying symmetry generators $X_{A(B)}$. Any superposition of these four states remains frozen under dynamics generated by Eq.~\eqref{eq:Heff}.
    }
    \label{fig:qubits}
\end{figure}

\textit{Model and frozen states.}---Our starting point is the $\mathsf{CZ}_p$ model of \cite{SHN}, whose Hamiltonian is given by
\begin{align}
\label{eq:CZP}
    H_{\mathsf{CZ}_{p}} = -J\sum_p \mathsf{CZ}_{p} - h\sum_i X_i,
\end{align}
where the sum in the first term runs over plaquettes $p$ of an $L\times L$ square lattice with periodic boundary conditions. 
$X_i$ and $Z_i$ are Pauli operators acting on qubit $i$. 
The potential term $\mathsf{CZ}_p=\prod_{\langle i,j\rangle\in p}\mathsf{CZ}_{ij}$ consists of a product of controlled-$Z$ gates $\mathsf{CZ}_{ij}=\frac{1}{2}(\mathbbm 1+Z_i+Z_j-Z_iZ_j)$ between nearest-neighbor pairs of qubits $i,j$ around the plaquette $p$.
The $\mathsf{CZ}_p$ model has a $\mathbb Z_2\times\mathbb Z_2$ symmetry generated by $X_{A(B)}=\prod_{i\in A(B)}X_i$, where $A$ and $B$ are the two sublattices of the square lattice.
This leads to a natural dual description in terms of sublattice domain walls, which can be visualized as line segments crossing each plaquette, see Fig.~\ref{fig:qubits}.
These domain walls must either form closed loops or large noncontractible loops that fully traverse the system.
In a given $z$-basis configuration, the potential term assigns an energy penalty $+J$ to each plaquette hosting two domain walls, corresponding to an intersection of two loops.
The kinetic term induces local fluctuations of these loops, including dynamical crossings and uncrossings.

HSF emerges in the limit $J\to\infty$, where the number of loop intersections, $N_{\rm int}=\sum_p(\mathbbm 1-\mathsf{CZ}_p)/2$, becomes a conserved quantity.
The effective model in this limit reads
\begin{align}
\label{eq:Heff}
    H_{\rm eff}=-h\sum_i X_i(P^{0000}_i+P^{1111}_{i}),
\end{align}
where $P^{0000}_i$ and $P^{1111}_i$ project $i$'s four nearest-neighbor sites onto $\ket 0$ or $\ket{1}$, respectively.
This Hamiltonian produces local fluctuations of loops so long as no new loop crossings are generated.
There is a set of product states in which no spin is flippable under $H_{\rm eff}$---that is, in which no qubit has all of its neighbors in the $\ket 0$ or $\ket 1$ state.
These product states can be visualized in the dual picture as close-packed configurations of non-contractible loops---for a few examples, see Fig.~\ref{fig:qubits}.
The number of such frozen states grows exponentially with system size, the exact number being $N_{\rm f}=2^{L+2}-8$.

These frozen states constitute good classical memories in the sense that they remain frozen in the presence of arbitrary perturbations.
To see this, we first note that these states remain frozen under Eq.~\eqref{eq:CZP} to any finite order in perturbation theory in the limit $h\ll J$.
The frozen states reside in the lowest-energy eigenspace of the potential term in Eq.~\eqref{eq:CZP}.
Within degenerate perturbation theory, the leading order corrections to these states arise from processes that mix states within this eigenspace.
The lowest-order process that accomplishes this requires one to flip $O(L)$ qubits along a noncontractible loop, which either maps one frozen state to another or maps a frozen state into a configuration with flippable spins.
Thus, in the thermodynamic limit, the frozen states remain frozen to arbitrarily late times.
In fact, the same argument applies to any $k$-local perturbation with coupling strength $\lambda \ll J$, provided $k/L\to 0$ as $L\to\infty$.

\textit{Encoded qubits.}---To convert the classical memory associated with the frozen states into a quantum one, we leverage the $\mathbb Z_2\times\mathbb Z_2$ symmetry of the model~\cite{Iadecola25}. 
More concretely, let us assign a label $\alpha=1,\dots,N_{\rm f}$ to each of the frozen states defined above. 
Then, for any frozen state $\ket{\alpha}$, we can define two sets of logical operators labeled by a sublattice index $s=A,B$:
\begin{align}
\label{eq:algebra}
\begin{split}
\mathcal I_{\alpha,s}=P_\alpha+X_s P_\alpha X_s &,\indent \mathcal Z_{\alpha,s}=P_\alpha-X_s P_\alpha X_s,\\
\mathcal X_{\alpha,s}=\{P_\alpha,X_s\}&,\indent i\mathcal Y_{\alpha,s}=[P_\alpha,X_s]
\end{split}
\end{align}
where $P_\alpha=\ket{\alpha}\bra{\alpha}$.
These operators satisfy $\mathcal X_{\alpha,s}^2=\mathcal Y_{\alpha,s}^2=\mathcal Z_{\alpha,s}^2=\mathcal I_{\alpha,s}$ and generate a Pauli algebra $[\mathcal X_{\alpha,s},\mathcal Y_{\alpha,s}]=2i\, \mathcal Z_{\alpha,s}$, etc.
Each copy of the Pauli algebra implies the existence of one encoded qubit.
More concretely, if all the operators in Eq.~\eqref{eq:algebra} commute with the Hamiltonian $H$, and the frozen state $\ket{\alpha}$ has energy $E_\alpha$, then the states $X_A\ket{\alpha},X_B\ket{\alpha},$ and $X_AX_B\ket{\alpha}$ all have energy $E_\alpha$ as well.
We can therefore define logical $z$-basis states (see Fig.~\ref{fig:qubits})
\begin{align}
\label{eq:logical state def}
\ket{\alpha;\sigma_A\sigma_B}=X_A^{\sigma_A}X_B^{\sigma_B}\ket{\alpha},
\end{align}
where $\sigma_A,\sigma_B\in\{0,1\}$.
Any superposition of these four states cannot decohere, and so this collection of states encodes two logical qubits.
The total number of logical qubits encoded in this way is $N_{\rm q}=N_{\rm f}/2$: every frozen state $\ket{\alpha}$ is connected to three others by the $\mathbb Z_2\times\mathbb Z_2$ generators $X_{A,B}$, and each such collection of states encodes two qubits.

For which Hamiltonians is the above construction valid?
By construction, it is valid for $H=H_{\rm eff}$ [Eq.~\eqref{eq:Heff}], under which all frozen states have energy zero.
More generally, it is exactly valid for any $H$ with $\mathbb Z_2\times \mathbb Z_2$ symmetry and for which the states $\{\ket{\alpha}\}^{N_{\rm f}}_{\alpha=1}$ are eigenstates.
The latter condition is highly fine tuned.
However, it can be made less restrictive by appealing to the asymptotic stability of the frozen states under Hamiltonians of the form $H=H_{\rm eff}+\lambda V$ where $\lambda\ll h$ and $V$ is $k$-local with $k=O(1)$.
For such a Hamiltonian, the frozen states of $H_{\rm eff}$ remain adiabatically connected via Schrieffer Wolff transformation to states that are frozen in the thermodynamic limit. 
Consequently, if we further require that the perturbation $V$ be symmetric under $\mathbb Z_2\times \mathbb Z_2$~\footnote{Note that the $\mathbb Z_2$ subgroup generated by $X=X_AX_B$ is itself sufficient to protect exponentially many qubits.},
then there will exist a set of encoded qubits in one to one correspondence with the ones discussed above which will be stable for times exponentially long in $L$. 
In the thermodynamic limit, these exponentially many qubits are stable to arbitrary symmetric perturbations of $H_{\rm eff}$. 
In terms of the original $\mathsf{CZ}_p$ model [Eq.~\eqref{eq:CZP}], which has $\mathbb Z_2\times \mathbb Z_2$ symmetry at the microscopic level, they are stable in the thermodynamic limit for $h\ll J$, and are additionally stable to the same class of perturbations $\lambda V$ with $\lambda \ll h$.

\begin{figure}[t!]
    \centering
    \includegraphics[width=0.7\columnwidth]{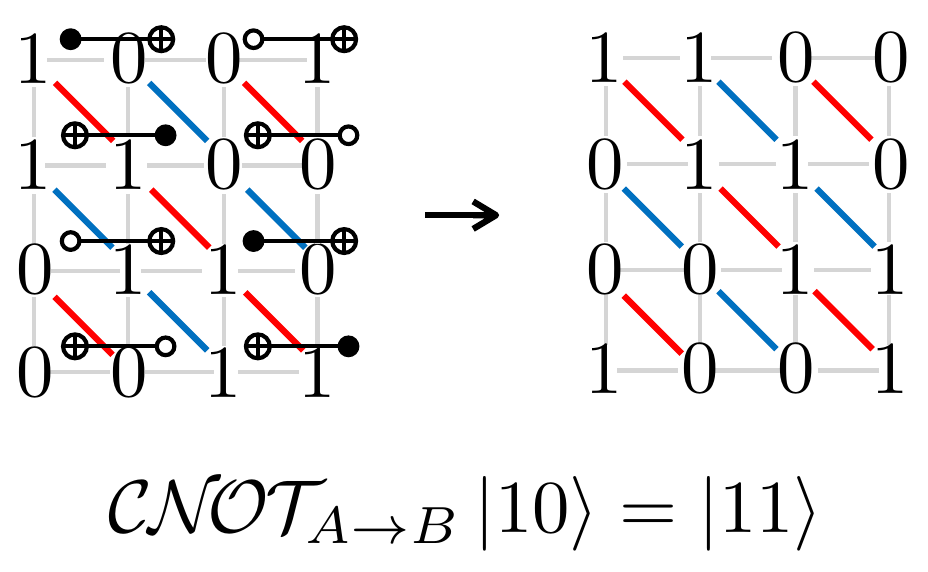}
    \caption{
    Schematic depiction of the logical $\mathsf{CNOT}$ gate defined in Eq.~\eqref{eq:CNOT}. This transversal logical $\mathsf{CNOT}$ gate is defined relative to the set of frozen states depicted in Fig.~\ref{fig:qubits}.
    }
    \label{fig:cnot}
\end{figure}

\textit{Logical gates.}---An appealing feature of the above construction based on $\mathbb Z_2\times\mathbb Z_2$ symmetry is that the logical qubit pairs defined by Eq.~\eqref{eq:algebra} admit a universal set of transversal logical gates. 
To construct these gates, it is convenient to write the frozen state $\ket{\alpha}=\bigotimes_{i}\ket{\sigma^\alpha_i}$.
Logical single-qubit rotations can then be constructed for any $\alpha$ and $s=A,B$ using the following logical $x$ and $z$ rotations:
\begin{align}
\begin{split}
    \mathcal R^{(\alpha,s)}_x(\theta) &= \prod_{j\in s} \exp\left(-i\frac{\theta}{2}X_j\right)\\
    \mathcal R^{(\alpha,s)}_z(\phi) &= \prod_{j\in s} \exp\left[-i(-1)^{\sigma^\alpha_j}\frac{\phi}{2N_s}Z_j\right],
\end{split}
\end{align}
where $N_s$ is the number of sites in sublattice $s$.

A transversal logical $\mathsf{CNOT}$ gate can also be constructed.
Let $\mathsf{CNOT}_{i\to j}$ denote a $\mathsf{CNOT}$ gate between physical qubits $i$ and $j$ with $i$ as the control.
Then the logical $\mathsf{CNOT}$ gate can be written as
\begin{align}
\label{eq:CNOT}
    \CNOT^{(\alpha)}_{A\to B} = \prod_{i\in A}X^{\sigma^\alpha_i}_i\mathsf{CNOT}_{i\to n(i)}X^{\sigma^\alpha_i}_i,
\end{align}
where $n(i)$ is the nearest neighbor to site $i$'s right (left) in odd (even) rows of the square lattice (for an example, see Fig.~\ref{fig:cnot}).
Dressing the physical $\mathsf{CNOT}$ gates with $X$ gates conditioned on the $A$ sublattice configuration ensures that the correct control logic is used for the logical $\mathsf{CNOT}$ gate~\footnote{Note that the frozen state $\ket{\alpha}$ is defined such that the $A$ logical qubit is in the $0$ state, see Eq.~\eqref{eq:logical state def}.}.

While this construction yields a set of logical qubits with a universal transversal gate set, it \textit{does not} yield a quantum error correcting code with such a gate set; this would violate the Eastin-Knill theorem~\cite{Eastin09}.
The reason we do not obtain an error correcting code is that, while $X$ errors on the physical qubits are detectable, $Z$ errors are not.
$X$ errors can be detected by measuring the plaquette stabilizers $Z_p = -\prod_{i\in p} Z_i$.
If $Z_p$ is measured with outcome $+1$, there is a single domain wall crossing the plaquette $p$; otherwise, there are either zero or two domain walls.
Both of the latter configurations are disallowed in the frozen state manifold.
However, a single physical $Z$ error can flip the eigenvalue of a logical $\mathcal X$ operator.
This results in a situation similar to a repetition code, which can be used to construct a logical qubit but not an error correcting code.
Like with the repetition code, one could imagine concatenating two copies of this frozen-state code in different bases to yield a Krylov-space analog of Shor's code.
This would allow the detection of physical $Z$ errors but spoil the transversal logical gate set.
Nevertheless, the resulting code may be of intrinsic interest as a many-body system, and we leave its study for future work.

\textit{Generalizations.}---While the discussion above has focused on frozen states, the construction of Ref.~\cite{Iadecola25} allows for encoding qubits in Krylov sectors of size greater than one. 
For example, in a model with a $\mathbb Z_2$ global symmetry, every Krylov sector that is \textit{not} invariant under the symmetry comes with a symmetry-related partner whose entire eigenspectrum is degenerate with that of the original sector.
While the model \eqref{eq:Heff} does not support topologically stable Krylov sectors of dimension greater than one, there is a related class of models, defined in Ref.~\cite{SNH}, that does.
As we show below, generalizing to this class of models allows one to encode \textit{qudits} in Krylov sectors of arbitrary size, provided they are not invariant under the protecting symmetry.

A simple illustration is provided by the {\it quad-flip model}, in which $m \ge3$-state clock spins are placed on every link of a square lattice. 
The local degrees of freedom are defined by the onsite clock and shift operators $Z_j$ and $X_j$, respectively, satisfying $X_j^m=Z_j^m=1$, $X_j^{\dag} = X_j^{m-1}$, $Z_j^{\dag} = Z_j^{m-1}$ and $Z_jX_j = e^{i2\pi/m} X_jZ_j$.
Denoting by $P^{(\kappa)}_{j}$ the projector onto a particular clock state $\ket{\kappa}$ ($\kappa=0,\dots,m-1$) on site $j$, one can define the ``flux density operator"
\begin{equation}
N_{\kappa, \ell} = \sum_{j \in \ell} (-1)^j P^{(\kappa)}_{j} 
\end{equation}
which measures the $\kappa$ flux through a loop $\ell$, with path-ordered $j$ and arbitrary but fixed origin. The elementary constraint is then that ``flux density" must vanish around every elementary plaquette. This can be interpreted as a loop model with $m$ species of colored loops and a non-intersection constraint between loops of different color. The elementary dynamics involves flipping the color of an elementary loop---i.e., if and only if all four links around an elementary plaquette are in the same state $|\kappa\rangle$, they can all be simultaneously flipped to $|\kappa'\rangle$. Now, the ordered sequence of non-contractible loops in non-repeating color is a topological invariant of the system, and fragments the Hilbert space into Krylov sectors labeled by this topological invariant. 

The quad-flip model has a global $\mathbb Z_m$ clock symmetry $X=\prod_{j}X_j$ that cycles the local clock variables $\ket{\kappa}\mapsto \ket{\kappa+1\mod m}$ on each site. 
When $m$ is prime, the Krylov sectors that transform nontrivially under $X$ must come in $m$-fold degenerate multiplets protected by the global clock symmetry.
When $m$ is not prime, the non-symmetric Krylov sectors must organize into multiplets whose dimension is a divisor of $m$; we henceforth assume $m$ is prime without loss of generality~\footnote{Note that, for composite $m$, the ensuing derivation goes through upon replacing $m$ by the multiplet size $n<m$.}.
Letting $P^{(0)}_\alpha$ denote the projector onto a non-symmetric Krylov sector, we are then guaranteed a set of symmetry-related sectors with projectors 
\begin{align}
P^{(\kappa)}_\alpha=X^{-\kappa} P^{(0)}_\alpha X^\kappa\indent (\kappa=0,\dots,m-1).
\end{align}
There are in general exponentially many such degenerate multiplets of Krylov sectors, which we differentiate using the label $\alpha$.

We can use this degeneracy to non-locally encode an $m$-state qudit with associated logical operators, working analogously to Eq.~\eqref{eq:algebra}. Specifically, given the orthogonal projectors $P^{(\kappa)}_\alpha$ for fixed $\alpha$, we can define logical operators
\begin{align}
\mathcal{I} \!=\! \sum_{\kappa = 0}^{m-1} P^{(\kappa)}_{\alpha},\ \mathcal{Z} \!=\! \sum_{\kappa = 0}^{m-1} e^{i \frac{2 \pi \kappa}{m}} P^{(\kappa)}_{\alpha},\ \mathcal{X} \!=\! \sum_{\kappa = 0}^{m-1} P^{(\kappa)}_{\alpha}X,
\end{align}
which obey $\mathcal Z^m=\mathcal X^m=\mathcal I$ and $\mathcal Z\mathcal X = e^{i2\pi/m} \mathcal X\mathcal Z$.
We reiterate that, in general, the projectors $P^{(\kappa)}_{\alpha}$ define Krylov subsectors of dimension larger than one. 
One consequence of this is that fixing the state of the logical qudit will {\it not} fix the state of a given physical qudit. 
This will make transversal application of logical gates difficult. 
However, it might mean that the encoding admits an error correction protocol. 
We leave investigation of this possibility to future work. 

It should also be noted that thus far our discussion has focused on models exhibiting {\it classical fragmentation}, i.e. HSF that manifests in an unentangled (product-state) basis. One can also have {\it quantum fragmentation}~\cite{moudgalya2022hilbert}, wherein the HSF only manifests in an entangled basis. 
The symmetry enrichment approach adopted here remains applicable in principle to quantum fragmented models and may be used to encode quantum information in such models.
However, insofar as single physical qubits do not have a well-defined state in the frozen eigenstates of a quantum fragmented model, it will likely not be possible to transversally apply logical gates. 
Whether this cost (inability to transversally apply logical gates) comes with a corresponding benefit (like existence of an error correcting encoding) is not presently clear to us, and would be a worthwhile problem for future investigation. 

\textit{Conclusion.}---We have discussed how combining ideas of topological HSF and symmetry enrichment can lead to exponentially many encoded qubits, which are topologically robust to arbitrary perturbations respecting a single discrete symmetry. We have illustrated this explicitly for the $\mathsf{CZ}_p$ model from Ref.~\cite{stephen2022ergodicity}, which we have further shown admits of transversal application of logical gates on the encoded qubits. The encoding is not robust to perturbations that break the protecting symmetry. We have discussed implications for more general encodings based on generalizations of topological fragmentation and quantum fragmentation. We hope this will open an interesting new path forward using exotic many body quantum dynamics to design new qubit encodings with useful properties. 

\begin{acknowledgments}
T.I. acknowledges support from the National Science Foundation under Grant Number~DMR-2143635. R.N. acknowledges support from the U.S. Department of
Energy, Office of Science, Basic Energy Sciences under Award
No. DE-SC0021346. 
This work was performed in part at the Kavli Institute for Theoretical Physics, which is supported by National Science Foundation grant PHY-2309135.
\end{acknowledgments}

\bibliography{refs}

\begin{thebibliography}{30}%
\makeatletter
\providecommand \@ifxundefined [1]{%
 \@ifx{#1\undefined}
}%
\providecommand \@ifnum [1]{%
 \ifnum #1\expandafter \@firstoftwo
 \else \expandafter \@secondoftwo
 \fi
}%
\providecommand \@ifx [1]{%
 \ifx #1\expandafter \@firstoftwo
 \else \expandafter \@secondoftwo
 \fi
}%
\providecommand \natexlab [1]{#1}%
\providecommand \enquote  [1]{``#1''}%
\providecommand \bibnamefont  [1]{#1}%
\providecommand \bibfnamefont [1]{#1}%
\providecommand \citenamefont [1]{#1}%
\providecommand \href@noop [0]{\@secondoftwo}%
\providecommand \href [0]{\begingroup \@sanitize@url \@href}%
\providecommand \@href[1]{\@@startlink{#1}\@@href}%
\providecommand \@@href[1]{\endgroup#1\@@endlink}%
\providecommand \@sanitize@url [0]{\catcode `\\12\catcode `\$12\catcode `\&12\catcode `\#12\catcode `\^12\catcode `\_12\catcode `\%12\relax}%
\providecommand \@@startlink[1]{}%
\providecommand \@@endlink[0]{}%
\providecommand \url  [0]{\begingroup\@sanitize@url \@url }%
\providecommand \@url [1]{\endgroup\@href {#1}{\urlprefix }}%
\providecommand \urlprefix  [0]{URL }%
\providecommand \Eprint [0]{\href }%
\providecommand \doibase [0]{https://doi.org/}%
\providecommand \selectlanguage [0]{\@gobble}%
\providecommand \bibinfo  [0]{\@secondoftwo}%
\providecommand \bibfield  [0]{\@secondoftwo}%
\providecommand \translation [1]{[#1]}%
\providecommand \BibitemOpen [0]{}%
\providecommand \bibitemStop [0]{}%
\providecommand \bibitemNoStop [0]{.\EOS\space}%
\providecommand \EOS [0]{\spacefactor3000\relax}%
\providecommand \BibitemShut  [1]{\csname bibitem#1\endcsname}%
\let\auto@bib@innerbib\@empty
\bibitem [{\citenamefont {Baxter}(2016)}]{Baxter}%
  \BibitemOpen
  \bibfield  {author} {\bibinfo {author} {\bibfnamefont {R.~J.}\ \bibnamefont {Baxter}},\ }\href@noop {} {\emph {\bibinfo {title} {Exactly solved models in statistical mechanics}}}\ (\bibinfo  {publisher} {Elsevier},\ \bibinfo {year} {2016})\BibitemShut {NoStop}%
\bibitem [{\citenamefont {Nandkishore}\ and\ \citenamefont {Huse}(2015)}]{MBLARCMP}%
  \BibitemOpen
  \bibfield  {author} {\bibinfo {author} {\bibfnamefont {R.}~\bibnamefont {Nandkishore}}\ and\ \bibinfo {author} {\bibfnamefont {D.~A.}\ \bibnamefont {Huse}},\ }\bibfield  {title} {\bibinfo {title} {Many-body localization and thermalization in quantum statistical mechanics},\ }\href {https://doi.org/10.1146/annurev-conmatphys-031214-014726} {\bibfield  {journal} {\bibinfo  {journal} {Annu. Rev. Condens. Matter Phys.}\ }\textbf {\bibinfo {volume} {6}},\ \bibinfo {pages} {15} (\bibinfo {year} {2015})}\BibitemShut {NoStop}%
\bibitem [{\citenamefont {Abanin}\ \emph {et~al.}(2019)\citenamefont {Abanin}, \citenamefont {Altman}, \citenamefont {Bloch},\ and\ \citenamefont {Serbyn}}]{MBLRMP}%
  \BibitemOpen
  \bibfield  {author} {\bibinfo {author} {\bibfnamefont {D.~A.}\ \bibnamefont {Abanin}}, \bibinfo {author} {\bibfnamefont {E.}~\bibnamefont {Altman}}, \bibinfo {author} {\bibfnamefont {I.}~\bibnamefont {Bloch}},\ and\ \bibinfo {author} {\bibfnamefont {M.}~\bibnamefont {Serbyn}},\ }\bibfield  {title} {\bibinfo {title} {Colloquium: Many-body localization, thermalization, and entanglement},\ }\href {https://doi.org/10.1103/RevModPhys.91.021001} {\bibfield  {journal} {\bibinfo  {journal} {Rev. Mod. Phys.}\ }\textbf {\bibinfo {volume} {91}},\ \bibinfo {pages} {021001} (\bibinfo {year} {2019})}\BibitemShut {NoStop}%
\bibitem [{\citenamefont {Serbyn}\ \emph {et~al.}(2021)\citenamefont {Serbyn}, \citenamefont {Abanin},\ and\ \citenamefont {Papić}}]{Serbyn21}%
  \BibitemOpen
  \bibfield  {author} {\bibinfo {author} {\bibfnamefont {M.}~\bibnamefont {Serbyn}}, \bibinfo {author} {\bibfnamefont {D.~A.}\ \bibnamefont {Abanin}},\ and\ \bibinfo {author} {\bibfnamefont {Z.}~\bibnamefont {Papić}},\ }\bibfield  {title} {\bibinfo {title} {Quantum many-body scars and weak breaking of ergodicity},\ }\href {https://doi.org/10.1038/s41567-021-01230-2} {\bibfield  {journal} {\bibinfo  {journal} {Nature Physics}\ }\textbf {\bibinfo {volume} {17}},\ \bibinfo {pages} {675–685} (\bibinfo {year} {2021})}\BibitemShut {NoStop}%
\bibitem [{\citenamefont {Moudgalya}\ \emph {et~al.}(2022{\natexlab{a}})\citenamefont {Moudgalya}, \citenamefont {Bernevig},\ and\ \citenamefont {Regnault}}]{Moudgalya22a}%
  \BibitemOpen
  \bibfield  {author} {\bibinfo {author} {\bibfnamefont {S.}~\bibnamefont {Moudgalya}}, \bibinfo {author} {\bibfnamefont {B.~A.}\ \bibnamefont {Bernevig}},\ and\ \bibinfo {author} {\bibfnamefont {N.}~\bibnamefont {Regnault}},\ }\bibfield  {title} {\bibinfo {title} {Quantum many-body scars and hilbert space fragmentation: a review of exact results},\ }\href {https://doi.org/10.1088/1361-6633/ac73a0} {\bibfield  {journal} {\bibinfo  {journal} {Reports on Progress in Physics}\ }\textbf {\bibinfo {volume} {85}},\ \bibinfo {pages} {086501} (\bibinfo {year} {2022}{\natexlab{a}})}\BibitemShut {NoStop}%
\bibitem [{\citenamefont {Chandran}\ \emph {et~al.}(2023)\citenamefont {Chandran}, \citenamefont {Iadecola}, \citenamefont {Khemani},\ and\ \citenamefont {Moessner}}]{Chandran23}%
  \BibitemOpen
  \bibfield  {author} {\bibinfo {author} {\bibfnamefont {A.}~\bibnamefont {Chandran}}, \bibinfo {author} {\bibfnamefont {T.}~\bibnamefont {Iadecola}}, \bibinfo {author} {\bibfnamefont {V.}~\bibnamefont {Khemani}},\ and\ \bibinfo {author} {\bibfnamefont {R.}~\bibnamefont {Moessner}},\ }\bibfield  {title} {\bibinfo {title} {Quantum many-body scars: A quasiparticle perspective},\ }\href {https://doi.org/10.1146/annurev-conmatphys-031620-101617} {\bibfield  {journal} {\bibinfo  {journal} {Annual Review of Condensed Matter Physics}\ }\textbf {\bibinfo {volume} {14}},\ \bibinfo {pages} {443–469} (\bibinfo {year} {2023})}\BibitemShut {NoStop}%
\bibitem [{\citenamefont {Khemani}\ \emph {et~al.}(2020)\citenamefont {Khemani}, \citenamefont {Hermele},\ and\ \citenamefont {Nandkishore}}]{KHN}%
  \BibitemOpen
  \bibfield  {author} {\bibinfo {author} {\bibfnamefont {V.}~\bibnamefont {Khemani}}, \bibinfo {author} {\bibfnamefont {M.}~\bibnamefont {Hermele}},\ and\ \bibinfo {author} {\bibfnamefont {R.}~\bibnamefont {Nandkishore}},\ }\bibfield  {title} {\bibinfo {title} {Localization from {H}ilbert space shattering: {F}rom theory to physical realizations},\ }\href {https://doi.org/10.1103/PhysRevB.101.174204} {\bibfield  {journal} {\bibinfo  {journal} {Phys. Rev. B}\ }\textbf {\bibinfo {volume} {101}},\ \bibinfo {pages} {174204} (\bibinfo {year} {2020})}\BibitemShut {NoStop}%
\bibitem [{\citenamefont {Sala}\ \emph {et~al.}(2020)\citenamefont {Sala}, \citenamefont {Rakovszky}, \citenamefont {Verresen}, \citenamefont {Knap},\ and\ \citenamefont {Pollmann}}]{Sala20}%
  \BibitemOpen
  \bibfield  {author} {\bibinfo {author} {\bibfnamefont {P.}~\bibnamefont {Sala}}, \bibinfo {author} {\bibfnamefont {T.}~\bibnamefont {Rakovszky}}, \bibinfo {author} {\bibfnamefont {R.}~\bibnamefont {Verresen}}, \bibinfo {author} {\bibfnamefont {M.}~\bibnamefont {Knap}},\ and\ \bibinfo {author} {\bibfnamefont {F.}~\bibnamefont {Pollmann}},\ }\bibfield  {title} {\bibinfo {title} {Ergodicity breaking arising from hilbert space fragmentation in dipole-conserving hamiltonians},\ }\href {https://doi.org/10.1103/PhysRevX.10.011047} {\bibfield  {journal} {\bibinfo  {journal} {Phys. Rev. X}\ }\textbf {\bibinfo {volume} {10}},\ \bibinfo {pages} {011047} (\bibinfo {year} {2020})}\BibitemShut {NoStop}%
\bibitem [{\citenamefont {Moudgalya}\ \emph {et~al.}(2022{\natexlab{b}})\citenamefont {Moudgalya}, \citenamefont {Prem}, \citenamefont {Nandkishore}, \citenamefont {Regnault},\ and\ \citenamefont {Bernevig}}]{Moudgalya2022Thermalization}%
  \BibitemOpen
  \bibfield  {author} {\bibinfo {author} {\bibfnamefont {S.}~\bibnamefont {Moudgalya}}, \bibinfo {author} {\bibfnamefont {A.}~\bibnamefont {Prem}}, \bibinfo {author} {\bibfnamefont {R.}~\bibnamefont {Nandkishore}}, \bibinfo {author} {\bibfnamefont {N.}~\bibnamefont {Regnault}},\ and\ \bibinfo {author} {\bibfnamefont {B.~A.}\ \bibnamefont {Bernevig}},\ }\bibfield  {title} {\bibinfo {title} {Thermalization and its absence within {K}rylov subspaces of a constrained {H}amiltonian},\ }in\ \href {https://doi.org/10.1142/9789811231711_0009} {\emph {\bibinfo {booktitle} {Memorial Volume for Shoucheng Zhang}}}\ (\bibinfo  {publisher} {World Scientific},\ \bibinfo {year} {2022})\ pp.\ \bibinfo {pages} {147--209}\BibitemShut {NoStop}%
\bibitem [{\citenamefont {Pai}\ \emph {et~al.}(2019)\citenamefont {Pai}, \citenamefont {Pretko},\ and\ \citenamefont {Nandkishore}}]{PPN}%
  \BibitemOpen
  \bibfield  {author} {\bibinfo {author} {\bibfnamefont {S.}~\bibnamefont {Pai}}, \bibinfo {author} {\bibfnamefont {M.}~\bibnamefont {Pretko}},\ and\ \bibinfo {author} {\bibfnamefont {R.~M.}\ \bibnamefont {Nandkishore}},\ }\bibfield  {title} {\bibinfo {title} {Localization in fractonic random circuits},\ }\href {https://doi.org/10.1103/PhysRevX.9.021003} {\bibfield  {journal} {\bibinfo  {journal} {Phys. Rev. X}\ }\textbf {\bibinfo {volume} {9}},\ \bibinfo {pages} {021003} (\bibinfo {year} {2019})}\BibitemShut {NoStop}%
\bibitem [{\citenamefont {Yoshinaga}\ \emph {et~al.}(2022)\citenamefont {Yoshinaga}, \citenamefont {Hakoshima}, \citenamefont {Imoto}, \citenamefont {Matsuzaki},\ and\ \citenamefont {Hamazaki}}]{Yoshinaga}%
  \BibitemOpen
  \bibfield  {author} {\bibinfo {author} {\bibfnamefont {A.}~\bibnamefont {Yoshinaga}}, \bibinfo {author} {\bibfnamefont {H.}~\bibnamefont {Hakoshima}}, \bibinfo {author} {\bibfnamefont {T.}~\bibnamefont {Imoto}}, \bibinfo {author} {\bibfnamefont {Y.}~\bibnamefont {Matsuzaki}},\ and\ \bibinfo {author} {\bibfnamefont {R.}~\bibnamefont {Hamazaki}},\ }\bibfield  {title} {\bibinfo {title} {Emergence of hilbert space fragmentation in ising models with a weak transverse field},\ }\href {https://doi.org/10.1103/PhysRevLett.129.090602} {\bibfield  {journal} {\bibinfo  {journal} {Phys. Rev. Lett.}\ }\textbf {\bibinfo {volume} {129}},\ \bibinfo {pages} {090602} (\bibinfo {year} {2022})}\BibitemShut {NoStop}%
\bibitem [{\citenamefont {Hart}\ and\ \citenamefont {Nandkishore}(2022)}]{HartIsing}%
  \BibitemOpen
  \bibfield  {author} {\bibinfo {author} {\bibfnamefont {O.}~\bibnamefont {Hart}}\ and\ \bibinfo {author} {\bibfnamefont {R.}~\bibnamefont {Nandkishore}},\ }\bibfield  {title} {\bibinfo {title} {Hilbert space shattering and dynamical freezing in the quantum ising model},\ }\href {https://doi.org/10.1103/PhysRevB.106.214426} {\bibfield  {journal} {\bibinfo  {journal} {Phys. Rev. B}\ }\textbf {\bibinfo {volume} {106}},\ \bibinfo {pages} {214426} (\bibinfo {year} {2022})}\BibitemShut {NoStop}%
\bibitem [{\citenamefont {Yang}\ \emph {et~al.}(2020)\citenamefont {Yang}, \citenamefont {Liu}, \citenamefont {Gorshkov},\ and\ \citenamefont {Iadecola}}]{Yang20}%
  \BibitemOpen
  \bibfield  {author} {\bibinfo {author} {\bibfnamefont {Z.-C.}\ \bibnamefont {Yang}}, \bibinfo {author} {\bibfnamefont {F.}~\bibnamefont {Liu}}, \bibinfo {author} {\bibfnamefont {A.~V.}\ \bibnamefont {Gorshkov}},\ and\ \bibinfo {author} {\bibfnamefont {T.}~\bibnamefont {Iadecola}},\ }\bibfield  {title} {\bibinfo {title} {Hilbert-space fragmentation from strict confinement},\ }\href {https://doi.org/10.1103/PhysRevLett.124.207602} {\bibfield  {journal} {\bibinfo  {journal} {Phys. Rev. Lett.}\ }\textbf {\bibinfo {volume} {124}},\ \bibinfo {pages} {207602} (\bibinfo {year} {2020})}\BibitemShut {NoStop}%
\bibitem [{\citenamefont {Stephen}\ \emph {et~al.}(2024{\natexlab{a}})\citenamefont {Stephen}, \citenamefont {Hart},\ and\ \citenamefont {Nandkishore}}]{stephen2022ergodicity}%
  \BibitemOpen
  \bibfield  {author} {\bibinfo {author} {\bibfnamefont {D.~T.}\ \bibnamefont {Stephen}}, \bibinfo {author} {\bibfnamefont {O.}~\bibnamefont {Hart}},\ and\ \bibinfo {author} {\bibfnamefont {R.~M.}\ \bibnamefont {Nandkishore}},\ }\bibfield  {title} {\bibinfo {title} {Ergodicity breaking provably robust to arbitrary perturbations},\ }\href {https://doi.org/10.1103/PhysRevLett.132.040401} {\bibfield  {journal} {\bibinfo  {journal} {Phys. Rev. Lett.}\ }\textbf {\bibinfo {volume} {132}},\ \bibinfo {pages} {040401} (\bibinfo {year} {2024}{\natexlab{a}})}\BibitemShut {NoStop}%
\bibitem [{\citenamefont {Stahl}\ \emph {et~al.}(2024)\citenamefont {Stahl}, \citenamefont {Nandkishore},\ and\ \citenamefont {Hart}}]{SNH}%
  \BibitemOpen
  \bibfield  {author} {\bibinfo {author} {\bibfnamefont {C.}~\bibnamefont {Stahl}}, \bibinfo {author} {\bibfnamefont {R.}~\bibnamefont {Nandkishore}},\ and\ \bibinfo {author} {\bibfnamefont {O.}~\bibnamefont {Hart}},\ }\bibfield  {title} {\bibinfo {title} {{Topologically stable ergodicity breaking from emergent higher-form symmetries in generalized quantum loop models}},\ }\href {https://doi.org/10.21468/SciPostPhys.16.3.068} {\bibfield  {journal} {\bibinfo  {journal} {SciPost Phys.}\ }\textbf {\bibinfo {volume} {16}},\ \bibinfo {pages} {068} (\bibinfo {year} {2024})}\BibitemShut {NoStop}%
\bibitem [{\citenamefont {Khudorozhkov}\ \emph {et~al.}(2025)\citenamefont {Khudorozhkov}, \citenamefont {Stahl}, \citenamefont {Hart},\ and\ \citenamefont {Nandkishore}}]{KSHN}%
  \BibitemOpen
  \bibfield  {author} {\bibinfo {author} {\bibfnamefont {A.}~\bibnamefont {Khudorozhkov}}, \bibinfo {author} {\bibfnamefont {C.}~\bibnamefont {Stahl}}, \bibinfo {author} {\bibfnamefont {O.}~\bibnamefont {Hart}},\ and\ \bibinfo {author} {\bibfnamefont {R.}~\bibnamefont {Nandkishore}},\ }\bibfield  {title} {\bibinfo {title} {Robust hilbert space fragmentation in group-valued loop models},\ }\href {https://doi.org/10.1103/PhysRevB.111.024310} {\bibfield  {journal} {\bibinfo  {journal} {Phys. Rev. B}\ }\textbf {\bibinfo {volume} {111}},\ \bibinfo {pages} {024310} (\bibinfo {year} {2025})}\BibitemShut {NoStop}%
\bibitem [{\citenamefont {Stahl}\ \emph {et~al.}(2025)\citenamefont {Stahl}, \citenamefont {Hart},\ and\ \citenamefont {Nandkishore}}]{robuster}%
  \BibitemOpen
  \bibfield  {author} {\bibinfo {author} {\bibfnamefont {C.}~\bibnamefont {Stahl}}, \bibinfo {author} {\bibfnamefont {O.}~\bibnamefont {Hart}},\ and\ \bibinfo {author} {\bibfnamefont {R.}~\bibnamefont {Nandkishore}},\ }\bibfield  {title} {\bibinfo {title} {Towards absolutely stable ergodicity breaking in two and three dimensions},\ }\href {https://doi.org/10.1103/PhysRevB.111.L020302} {\bibfield  {journal} {\bibinfo  {journal} {Phys. Rev. B}\ }\textbf {\bibinfo {volume} {111}},\ \bibinfo {pages} {L020302} (\bibinfo {year} {2025})}\BibitemShut {NoStop}%
\bibitem [{\citenamefont {Moudgalya}\ and\ \citenamefont {Motrunich}(2022)}]{moudgalya2022hilbert}%
  \BibitemOpen
  \bibfield  {author} {\bibinfo {author} {\bibfnamefont {S.}~\bibnamefont {Moudgalya}}\ and\ \bibinfo {author} {\bibfnamefont {O.~I.}\ \bibnamefont {Motrunich}},\ }\bibfield  {title} {\bibinfo {title} {Hilbert space fragmentation and commutant algebras},\ }\href {https://doi.org/10.1103/PhysRevX.12.011050} {\bibfield  {journal} {\bibinfo  {journal} {Physical Review X}\ }\textbf {\bibinfo {volume} {12}},\ \bibinfo {pages} {011050} (\bibinfo {year} {2022})}\BibitemShut {NoStop}%
\bibitem [{\citenamefont {Balasubramanian}\ \emph {et~al.}(2024)\citenamefont {Balasubramanian}, \citenamefont {Gopalakrishnan}, \citenamefont {Khudorozhkov},\ and\ \citenamefont {Lake}}]{balasubramanian2023glassy}%
  \BibitemOpen
  \bibfield  {author} {\bibinfo {author} {\bibfnamefont {S.}~\bibnamefont {Balasubramanian}}, \bibinfo {author} {\bibfnamefont {S.}~\bibnamefont {Gopalakrishnan}}, \bibinfo {author} {\bibfnamefont {A.}~\bibnamefont {Khudorozhkov}},\ and\ \bibinfo {author} {\bibfnamefont {E.}~\bibnamefont {Lake}},\ }\bibfield  {title} {\bibinfo {title} {Glassy word problems: Ultraslow relaxation, {H}ilbert space jamming, and computational complexity},\ }\href {https://doi.org/10.1103/PhysRevX.14.021034} {\bibfield  {journal} {\bibinfo  {journal} {Phys. Rev. X}\ }\textbf {\bibinfo {volume} {14}},\ \bibinfo {pages} {021034} (\bibinfo {year} {2024})}\BibitemShut {NoStop}%
\bibitem [{\citenamefont {Iadecola}(2025)}]{Iadecola25}%
  \BibitemOpen
  \bibfield  {author} {\bibinfo {author} {\bibfnamefont {T.}~\bibnamefont {Iadecola}},\ }\href@noop {} {\bibinfo {title} {Symmetry fragmentation}} (\bibinfo {year} {2025}),\ \Eprint {https://arxiv.org/abs/arXiv:2510.06333} {arXiv:2510.06333} \BibitemShut {NoStop}%
\bibitem [{\citenamefont {Singh}\ \emph {et~al.}(2021)\citenamefont {Singh}, \citenamefont {Ware}, \citenamefont {Vasseur},\ and\ \citenamefont {Friedman}}]{Singh21}%
  \BibitemOpen
  \bibfield  {author} {\bibinfo {author} {\bibfnamefont {H.}~\bibnamefont {Singh}}, \bibinfo {author} {\bibfnamefont {B.~A.}\ \bibnamefont {Ware}}, \bibinfo {author} {\bibfnamefont {R.}~\bibnamefont {Vasseur}},\ and\ \bibinfo {author} {\bibfnamefont {A.~J.}\ \bibnamefont {Friedman}},\ }\bibfield  {title} {\bibinfo {title} {Subdiffusion and many-body quantum chaos with kinetic constraints},\ }\href {https://doi.org/10.1103/PhysRevLett.127.230602} {\bibfield  {journal} {\bibinfo  {journal} {Phys. Rev. Lett.}\ }\textbf {\bibinfo {volume} {127}},\ \bibinfo {pages} {230602} (\bibinfo {year} {2021})}\BibitemShut {NoStop}%
\bibitem [{\citenamefont {Pozsgay}\ \emph {et~al.}(2021)\citenamefont {Pozsgay}, \citenamefont {Gombor}, \citenamefont {Hutsalyuk}, \citenamefont {Jiang}, \citenamefont {Pristy\'ak},\ and\ \citenamefont {Vernier}}]{Pozsgay21}%
  \BibitemOpen
  \bibfield  {author} {\bibinfo {author} {\bibfnamefont {B.}~\bibnamefont {Pozsgay}}, \bibinfo {author} {\bibfnamefont {T.}~\bibnamefont {Gombor}}, \bibinfo {author} {\bibfnamefont {A.}~\bibnamefont {Hutsalyuk}}, \bibinfo {author} {\bibfnamefont {Y.}~\bibnamefont {Jiang}}, \bibinfo {author} {\bibfnamefont {L.}~\bibnamefont {Pristy\'ak}},\ and\ \bibinfo {author} {\bibfnamefont {E.}~\bibnamefont {Vernier}},\ }\bibfield  {title} {\bibinfo {title} {Integrable spin chain with hilbert space fragmentation and solvable real-time dynamics},\ }\href {https://doi.org/10.1103/PhysRevE.104.044106} {\bibfield  {journal} {\bibinfo  {journal} {Phys. Rev. E}\ }\textbf {\bibinfo {volume} {104}},\ \bibinfo {pages} {044106} (\bibinfo {year} {2021})}\BibitemShut {NoStop}%
\bibitem [{\citenamefont {Zadnik}\ and\ \citenamefont {Fagotti}(2021)}]{Zadnik21a}%
  \BibitemOpen
  \bibfield  {author} {\bibinfo {author} {\bibfnamefont {L.}~\bibnamefont {Zadnik}}\ and\ \bibinfo {author} {\bibfnamefont {M.}~\bibnamefont {Fagotti}},\ }\bibfield  {title} {\bibinfo {title} {{The Folded Spin-1/2 XXZ Model: I. Diagonalisation, Jamming, and Ground State Properties}},\ }\href {https://doi.org/10.21468/SciPostPhysCore.4.2.010} {\bibfield  {journal} {\bibinfo  {journal} {SciPost Phys. Core}\ }\textbf {\bibinfo {volume} {4}},\ \bibinfo {pages} {010} (\bibinfo {year} {2021})}\BibitemShut {NoStop}%
\bibitem [{\citenamefont {Zadnik}\ \emph {et~al.}(2021)\citenamefont {Zadnik}, \citenamefont {Bidzhiev},\ and\ \citenamefont {Fagotti}}]{Zadnik21b}%
  \BibitemOpen
  \bibfield  {author} {\bibinfo {author} {\bibfnamefont {L.}~\bibnamefont {Zadnik}}, \bibinfo {author} {\bibfnamefont {K.}~\bibnamefont {Bidzhiev}},\ and\ \bibinfo {author} {\bibfnamefont {M.}~\bibnamefont {Fagotti}},\ }\bibfield  {title} {\bibinfo {title} {{The folded spin-1/2 XXZ model: II. Thermodynamics and hydrodynamics with a minimal set of charges}},\ }\href {https://doi.org/10.21468/SciPostPhys.10.5.099} {\bibfield  {journal} {\bibinfo  {journal} {SciPost Phys.}\ }\textbf {\bibinfo {volume} {10}},\ \bibinfo {pages} {099} (\bibinfo {year} {2021})}\BibitemShut {NoStop}%
\bibitem [{\citenamefont {Vuina}\ \emph {et~al.}(2025)\citenamefont {Vuina}, \citenamefont {Schäfer}, \citenamefont {Long},\ and\ \citenamefont {Chandran}}]{Vuina25}%
  \BibitemOpen
  \bibfield  {author} {\bibinfo {author} {\bibfnamefont {D.}~\bibnamefont {Vuina}}, \bibinfo {author} {\bibfnamefont {R.}~\bibnamefont {Schäfer}}, \bibinfo {author} {\bibfnamefont {D.~M.}\ \bibnamefont {Long}},\ and\ \bibinfo {author} {\bibfnamefont {A.}~\bibnamefont {Chandran}},\ }\href@noop {} {\bibinfo {title} {Probing {Hilbert} space fragmentation using controlled dephasing}} (\bibinfo {year} {2025}),\ \Eprint {https://arxiv.org/abs/arXiv:2506.13856} {arXiv:2506.13856} \BibitemShut {NoStop}%
\bibitem [{\citenamefont {Eastin}\ and\ \citenamefont {Knill}(2009)}]{Eastin09}%
  \BibitemOpen
  \bibfield  {author} {\bibinfo {author} {\bibfnamefont {B.}~\bibnamefont {Eastin}}\ and\ \bibinfo {author} {\bibfnamefont {E.}~\bibnamefont {Knill}},\ }\bibfield  {title} {\bibinfo {title} {Restrictions on transversal encoded quantum gate sets},\ }\href {https://doi.org/10.1103/PhysRevLett.102.110502} {\bibfield  {journal} {\bibinfo  {journal} {Phys. Rev. Lett.}\ }\textbf {\bibinfo {volume} {102}},\ \bibinfo {pages} {110502} (\bibinfo {year} {2009})}\BibitemShut {NoStop}%
\bibitem [{\citenamefont {Stephen}\ \emph {et~al.}(2024{\natexlab{b}})\citenamefont {Stephen}, \citenamefont {Hart},\ and\ \citenamefont {Nandkishore}}]{SHN}%
  \BibitemOpen
  \bibfield  {author} {\bibinfo {author} {\bibfnamefont {D.~T.}\ \bibnamefont {Stephen}}, \bibinfo {author} {\bibfnamefont {O.}~\bibnamefont {Hart}},\ and\ \bibinfo {author} {\bibfnamefont {R.~M.}\ \bibnamefont {Nandkishore}},\ }\bibfield  {title} {\bibinfo {title} {Ergodicity breaking provably robust to arbitrary perturbations},\ }\href {https://doi.org/10.1103/PhysRevLett.132.040401} {\bibfield  {journal} {\bibinfo  {journal} {Phys. Rev. Lett.}\ }\textbf {\bibinfo {volume} {132}},\ \bibinfo {pages} {040401} (\bibinfo {year} {2024}{\natexlab{b}})}\BibitemShut {NoStop}%
\bibitem [{Note1()}]{Note1}%
  \BibitemOpen
  \bibinfo {note} {Note that the $\protect \mathbb Z_2$ subgroup generated by $X=X_AX_B$ is itself sufficient to protect exponentially many qubits.}\BibitemShut {Stop}%
\bibitem [{Note2()}]{Note2}%
  \BibitemOpen
  \bibinfo {note} {Note that the frozen state $\mathinner {|{\alpha }\rangle }$ is defined such that the $A$ logical qubit is in the $0$ state, see Eq.~\protect \eqref {eq:logical state def}.}\BibitemShut {Stop}%
\bibitem [{Note3()}]{Note3}%
  \BibitemOpen
  \bibinfo {note} {Note that, for composite $m$, the ensuing derivation goes through upon replacing $m$ by the multiplet size $n<m$.}\BibitemShut {Stop}%
\end{thebibliography}%




\end{document}